\documentclass[aps,twocolumn,showpacs,prl,amsmath,amssymb,floatfix,superscriptaddress]{revtex4-1}
\usepackage{amsmath,amssymb,graphicx,color}
\usepackage{bm}
\usepackage{bbm}
\usepackage[caption=false]{subfig}
\usepackage[usenames,dvipsnames]{xcolor}
\usepackage[colorlinks=true,citecolor=Cerulean,linkcolor=RubineRed,urlcolor=Cerulean]{hyperref}
\usepackage[normalem]{ulem}
\usepackage{soul,xcolor}
\newcommand{\sandwich}[3]{\langle #1 \vert #2 \vert #3 \rangle}

\newcommand{\mr}[1]{\mathrm{#1}}

\newcommand{\Tr}{\ensuremath{\mathrm{Tr}\,}}
\newcommand{\avg}[1]{\ensuremath{\langle #1 \rangle}}
\newcommand{\ket}[1]{\ensuremath{\vert #1 \rangle}}
\newcommand{\bra}[1]{\ensuremath{\langle #1 \vert}}
\newcommand{\wich}[3]{\ensuremath{\langle #1 \vert #2 \vert #3\rangle}}
\usepackage{fixltx2e,amsmath}
\MakeRobust{\eqref}
\newcommand{\eqnref}[1]{Eq.~\eqref{#1}}
\newcommand{\figref}[1]{Fig.~\ref{#1}}
\usepackage{mathptmx} 

\begin{document}
\setstcolor{blue}
\title{Quantum Statistical Enhancement of the Collective Performance of Multiple Bosonic Engines}

\author{Gentaro Watanabe}\email{gentaro@zju.edu.cn}
\affiliation{Department of Physics and Zhejiang Institute of Modern Physics, Zhejiang University, Hangzhou, Zhejiang 310027, China}
\affiliation{Zhejiang Province Key Laboratory of Quantum Technology and Device, Zhejiang University, Hangzhou, Zhejiang 310027, China}
\author{B. Prasanna Venkatesh}\email{prasanna.b@iitgn.ac.in}
\affiliation{Indian Institute of Technology Gandhinagar, Gandhinagar, Gujarat 382355, India}
\author{Peter Talkner}
\affiliation{Institut f\"{u}r Physik, Universit\"{a}t Augsburg, Universit\"{a}tsstra\ss e 1, D-86135 Augsburg, Germany}
\author{Myung-Joong Hwang}
\affiliation{Division of Natural Sciences, Duke Kunshan University, No.~8 Duke Avenue, Kunshan, Jiangsu 215316, China}
\affiliation{Institute for Theoretical Physics, Ulm University, Albert-Einstein Allee 11, D-89081 Ulm, Germany}
\author{Adolfo del Campo}
\affiliation{Donostia International Physics Center,  E-20018 San Sebasti\'an, Spain}
\affiliation{IKERBASQUE, Basque Foundation for Science, E-48013 Bilbao, Spain}
\affiliation{Department of Physics, University of Massachusetts, Boston, Massachusetts 02125, USA}


\date{\today}

\begin{abstract}
We consider an ensemble of indistinguishable quantum machines and show that quantum statistical effects can give rise to  a genuine quantum enhancement of the collective thermodynamic performance. When multiple indistinguishable bosonic work resources are coupled to an external system, the internal energy change of the external system exhibits an enhancement arising from permutation symmetry in the ensemble, which is absent when the latter consists of distinguishable work resources.
\end{abstract}

\pacs{03.65.Ta, 05.30.-d, 05.40.-a, 05.70.Ln}
\maketitle

\textit{Introduction.---}
Technological advances have allowed us to miniaturize thermal machines to the nanoscale and beyond, where quantum effects can play an important role  \cite{QTroadmap18}. A paradigmatic instance  of a thermal machine is  a quantum heat engine (QHE). First conceived in the late 1950s, a QHE is  a quantum system that serves as the working fuel of a thermodynamic cycle \cite{scovil59,alicki79,kosloff84,bender00,quan07,kosloff14}. More recently, a synergy between technology and progress on the foundations of quantum thermodynamics \cite{Talkner07,Campisi11,Goold16,vinjanampathy16,strasberg17,alicki18}   has led to a surge of activity  on the study of quantum machines \cite{Scully03,Scully11,Abah12,Abah14,Rossnagel14,Rossnagel16,Brantut13,maslennikov19,Klatzow19},  consolidating it as an active area of research \cite{ghosh18}. A prominent challenge in this context  is the identification of situations where quantum effects govern and lead to an enhanced performance with no classical counterpart \cite{Scully11,jaramillo16,Yi17,Ding18}.
One strategy to identify such situations is to consider thermal machines composed of multiple components \cite{Sisman01,Saygin01,gong14, zheng15,binder15,hardal15,Beau16,jaramillo16,uzdin16,campaioli17,hardal18,niedenzu18,andolina19,gelbwaser19,Klatzow19,manatuly18,chen18,myers20} described by collective quantum states with mutual coherence. In this setting, the enhancement requires either interaction among the components or the performance of collective unitary operations on the constituents. Furthermore, the natural process of extracting work from a QHE by outcoupling it to drive another quantum system \cite{levy16,watanabe17} or even the process of storing work in a quantum system \cite{binder15,andolina19} can also lead to the manifestation of genuine quantum effects. 
In this Letter, we identify a third route, in which quantum statistics leads to a genuine quantum enhancement of the performance as a result of the statistical indistinguishability of the constituent work resources. Specifically, we consider multiple work resources, each composed of a single QHE with an individual piston \cite{Abah12,gelbwaser14}, coupled to a single external system and show that the internal energy change of the external system displays quantum enhancement when the QHEs are indistinguishable. We note that such a setting is fundamentally different from a single QHE with a working fluid consisting of multiple particles \cite{Kim11,zheng15,Beau16,jaramillo16,Bengtsson18,hardal18,Deng18,chen18}.
\begin{figure}[b!]
\centering
\includegraphics[width=0.93 \columnwidth]{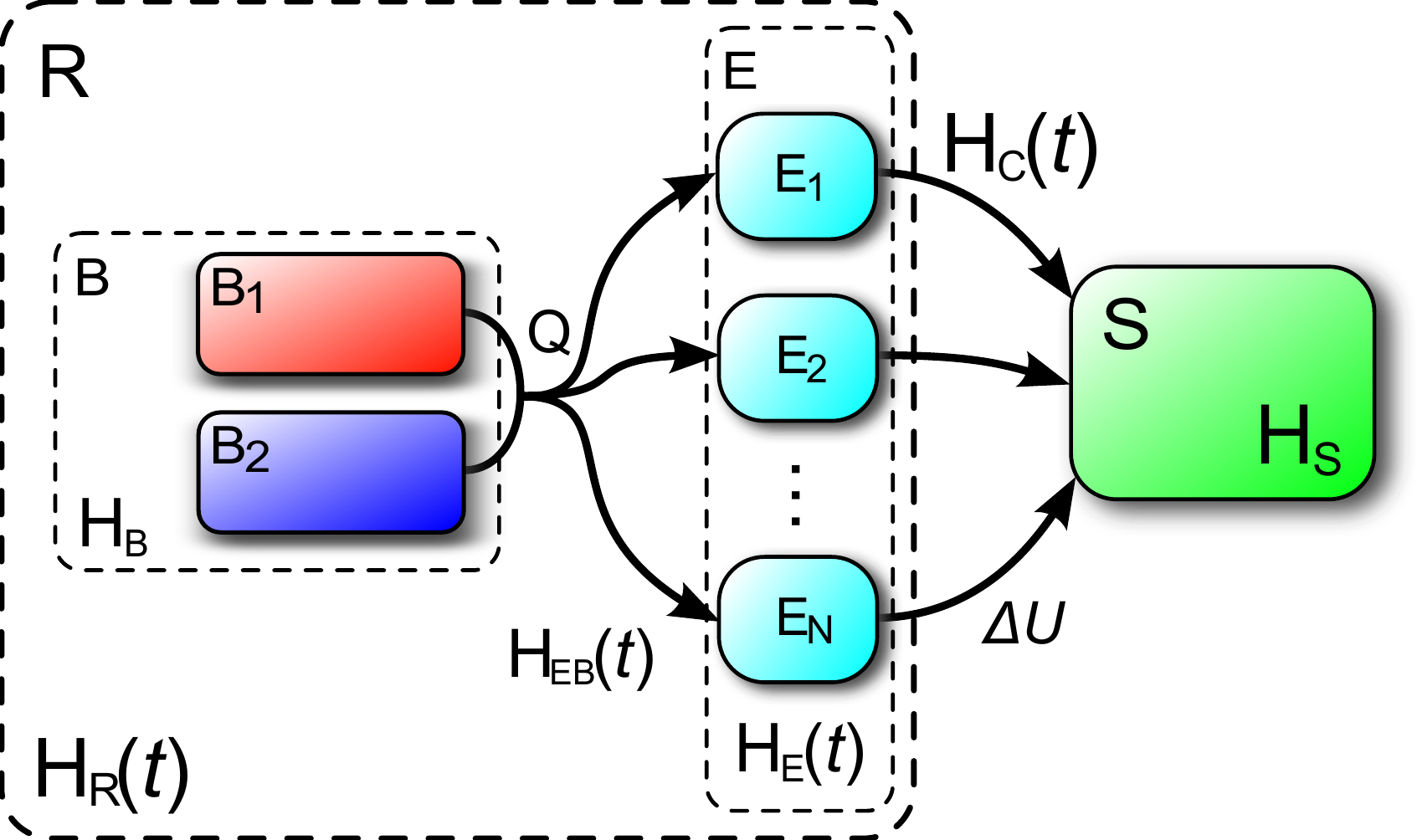}
\caption{
  {Schematic picture of the setup.}
  Multiple work resources collectively denoted by $R$ deliver energy $\Delta U$ to an external system $S$ through the coupling Hamiltonian $H_C$. If the work resources are $N$ quantum heat engines $E_1,\, \cdots,\, E_N$, all the engines and the heat baths collectively denoted by $E$ and $B$, respectively, are included in $R$.}
\label{fig:setup}
\end{figure}

\textit{Setup.---}
Consider a collective work resource $R$ made of $N$ heat engines $E_1$, $\cdots$, $E_N$ interacting with two heat baths ($B_1$ and $B_2$) and an external quantum system $S$ on which the work is performed. The coupling between $R$ and $S$ is solely established via the heat engines; see Fig.~\ref{fig:setup}. The global Hamiltonian of the whole system is the sum of that of the work resources, the external system, and the coupling $C$ between them:
\begin{align}
  H(t) = H_R(t) + H_C(t) + H_S\,,
\end{align}
where the external system is assumed to be time independent. If the work resources are QHEs, $H_R(t)$ collectively represents the Hamiltonian for $N$ engines and the two common baths. For simplicity, we consider the following form of the coupling Hamiltonian $H_C$:
\begin{align}
  H_C(t) = g_C(t)\, V_{R} \otimes V_{S}\,,\label{eq:hc_simp}
\end{align}
where $g_{C}(t)$ is a time-dependent coupling constant, and $V_{R}$ and $V_{S}$ are operators of the work resource and of the external system, respectively.
In the analytical treatment below, we assume a sufficiently weak coupling between the work resources and the external system justifying a perturbative treatment \cite{note:general}.

In this Letter, we characterize the work performed by the work resources on the system $S$ by its internal energy change $\Delta U$. A more in-depth discussion as to what extent $\Delta U$ represents work will be provided in the conclusions. $\Delta U$ is evaluated by energy measurements on the external system $S$ at the beginning and the end of the cycle at $t=0$ and $T$, respectively \cite{watanabe17}. For simplicity, we turn off the coupling $g_C(t)$ at $t=0$ and $T$, and, thus, $[H_S, H(t)]=0$ at these moments in time.
Consequently, measurements of the system energy $H_S$ at these two times do not affect the state of the work resources $R$.
The external system is initially prepared in its ground state $|0\rangle_S$, and the initial state $\rho_0$ of the total system is $\rho_0 = \rho^R_0 \otimes |0\rangle_S{_S}\langle 0|$ with $\rho^R_0$ being the initial state of the work resources.
\begin{figure}[t]
	\centering
	\includegraphics[width= \columnwidth]{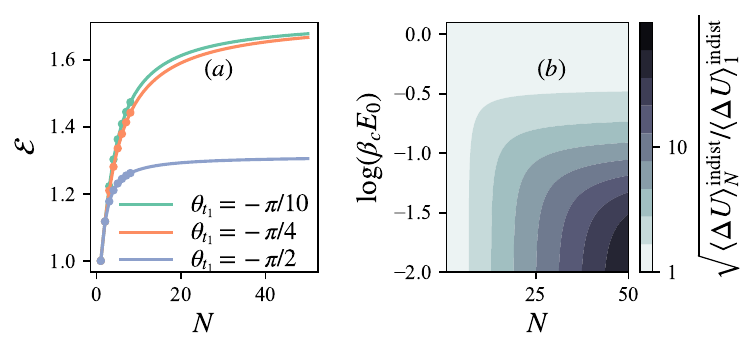}
	\caption{
		{Quantum-enhanced performance of multiple identical engines in the perturbative impulse coupling.}
		(a) Ratio $\mathcal{E} \equiv \avg{\Delta U}_N^{\rm indist}/\avg{\Delta U}_N^{\rm dist}$ for an outcoupling with an impulse of strength $g = 0.01$ to an external system. The atomic energy gap is driven with a linear speed of $v = 0.1 \Omega(0)^2$ over a total protocol time of $T = 20/\Omega(0)$. The impulse kick occurs at $t_1 = 0.35 T/2$, and $\theta_{t_1}$ is tuned by changing $\Delta = \{4.2 \Omega(0),1.4\Omega(0),0\}$ (different curves). The bath temperatures are given by $\beta_c E_0 = 2$ and $\beta_h E_{T/2} = 1/4$. Solid lines are from analytical expressions derived in the perturbative limit, and dots are from a numerical calculation with the external system given by a HO of frequency $\omega = 2\pi \times 0.05/T$. (b) $\avg{\Delta U}_N^{\rm indist}$ produced by indistinguishable particles when $\Delta = 0$ and $\beta_h E_{T/2} = 1/4$ as a function of $N$ and $\beta_c E_0$ for impulse-type coupling and other parameters as in (a). Regions with linear scaling in $N$ of the contour gradients correspond to quadratic scaling of $\avg{\Delta U}_N$ with $N$.}
	\label{fig:bose_identical_delta}
\end{figure}

\textit{Average energy delivered by outcoupled work resources.---}
We consider the average internal energy change of the system $S$, $\avg{\Delta U}_N$, caused by the coupling to the $N$ work resources. In the rotating frame with respect to $H_0(t) \equiv H_R(t) + H_S$, the propagator in the interaction picture is $U^{(I)}(t,0) = \mathcal{T} \exp{[-i \int_0^t H_C^{(I)}(t')\, dt']}$, where $\mathcal{T}$ is the time-ordering operator and $H_C^{(I)}(t) \equiv U_0^\dagger(t,0)\, H_C(t)\, U_0(t,0)$ with $U_0(t,0) \equiv \mathcal{T} \exp{[-i \int_0^t H_0(t')\, dt']}$. Regarding the coupling Hamiltonian $H_C$ in Eq.~(\ref{eq:hc_simp}), $H_C^{(I)}(t)$ reads $H_C^{(I)}(t) = g_C(t)\, V_R^{(I)}(t) \otimes V_S^{(I)}(t)$ with $V_R^{(I)}(t) \equiv U_R^\dagger(t,0)\, V_R\, U_R(t,0)$, $V_S^{(I)}(t) \equiv e^{iH_St}\, V_S\, e^{-iH_St}$, and $U_R(t,0) \equiv \mathcal{T} \exp{[-i \int_0^t H_R(t')\, dt']}$.

Setting the energy of the ground state $|0\rangle_S$ of the system to be zero without loss of generality, the average internal energy change is given by $\avg{\Delta U}_N =  \sum_{i \neq 0} \epsilon_i^S\, p_i$, where the probability $p_i$ for measuring the $i$th eigenvalue $\epsilon_i^S$ of $H_S$ as an outcome of the energy measurement at $t=T$ reads
\begin{align}
  p_i = \Tr_{R}\left[ {_S}\langle i| U^{(I)}(T,0)\, \rho_0\, {U^{(I)}}^\dagger(T,0) |i\rangle_S \right]\,.\label{eq:pi}
\end{align}
Here, $\Tr_R[\cdots]$ is the trace over the Hilbert space of the work resources and $|i\rangle_S$ is the $i$th eigenvector of $H_S$. To gain an analytical insight, we first resort to the weak coupling regime where $\int_0^T g_C(t)\, dt \ll 1$. In this limit, expanding the propagator to leading order as $U^{(I)}(T,0) \approx I - i \int_0^T dt\, g_C(t) V_R^{(I)}(t) \otimes V_S^{(I)}(t)$ in \eqnref{eq:pi}, the excitation probability of the system reduces to
\begin{align}
p_{i} \simeq \int_0^T dt \int_0^T dt^{\prime}\, &g_C(t)\, g_C(t^\prime)\,\,_S\wich{i}{V_S^{(I)}(t)}{0}_S \,_S\wich{0}{V_S^{(I)}(t^{\prime})}{i}_S \nonumber\\
& \langle V_R^{(I)}(t^{\prime})\, V_R^{(I)}(t) \rangle_{\rho_0^R}, \label{eq:probcc_gen}
\end{align}   
with $\avg{\cdots}_{\rho^R_0} \equiv \Tr_{R}[\cdots \rho^R_0]$. 

\textit{Quantum statistical enhancement.---}
To demonstrate the genuinely quantum mechanical advantage of indistinguishable bosons in comparison to distinguishable particles as the work resources, we consider $N$ QHEs, each performing an Otto cycle with the two lowest internal energy levels of a bosonic atom prepared in its center of mass (COM) ground state as a working fluid, i.e., the temperature $\beta_{\rm COM}^{-1}$ of the COM degrees of freedom is set to be zero. As sketched in Fig.~\ref{fig:setup}, the work resources $R$ contain these engines, together with the hot and cold heat baths.

The four strokes of the Otto cycle are performed as follows:
(0) \emph{Initial state.---} In the absence of the coupling to the external system $S$, $g_C(0)=0$, all two-level atoms are prepared in thermal equilibrium with the common cold bath at inverse temperature $\beta_c$. Thus, the initial reduced density matrix $\rho^E_0 \equiv \Tr_B{\,\rho^R_0}$ of the engine part is $\rho^E_0 = Z_{\beta_c}^{-1}\, \exp{[-\beta_c H_E(0)]}$ with $Z_{\beta_c} \equiv \Tr_E{\, \exp{[-\beta_c H_E(0)]}}$, where $\Tr_{E}[\cdots]$ and $\Tr_{B}[\cdots]$ are the trace over the Hilbert space of the engines and that of the baths, respectively. The baths are assumed to be time independent and in the canonical state of $H_B$ throughout the cycle.
(1) \emph{Isentropic compression.---} From $0 < t < T/2$, all the engines are decoupled from the baths, $H_{EB}=0$, and the level distance of all the two-level atoms is slowly increased in the same manner.
(2) \emph{Hot isochore.---} At $t=T/2$, setting $g_C=0$, all the two-level atoms are brought into weak contact with a common hot bath and thermalized at inverse temperature $\beta_h$.
At the end of this process, the state of the engine is given by $\rho^E_{T/2} =  \Tr_B{\,\rho^R_{T/2}} = Z_{\beta_h}^{-1}\, \exp{[-\beta_h H_E(T/2)]}$ with $Z_{\beta_h} \equiv \Tr_E{\, \exp{[-\beta_h H_E(T/2)]}}$.
(3) \emph{Isentropic expansion.---} From $T/2 < t < T$, all engines are decoupled from the baths, $H_{EB}=0$, and the energy separation of each two-level atoms is decreased slowly in the same way.
(4) \emph{Cold isochore.---} At $t=T$, setting $g_C=0$, and all the two-level atoms are brought into contact with the common cold bath again and quickly return to the initial state.

First, we focus on the case of indistinguishable atoms. We choose $V_R=2S_x$ in the coupling Hamiltonian (\ref{eq:hc_simp}):
\begin{align}
  H_C(t) = g_C(t)\, 2 S_x \otimes V_S\,,\label{eq:hc_sx}
\end{align}
with $S_x \equiv (a^\dagger b + b^\dagger a)/2$, where $a^\dagger$ and $a$ are creation and annihilation operators of the ground-state atoms in the lowest COM level, and $b^\dagger$ and $b$ are those of the excited-state atoms, respectively. While we keep the external system general in this discussion, we note that if the external system is a harmonic oscillator (HO) and $V_S = c^\dagger + c$ with $c$ ($c^{\dagger}$) denoting the annihilation (creation) operator of the HO, Eq.~(\ref{eq:hc_sx}) reduces to the standard dipole coupling between an ensemble of atoms and a single-mode HO. In order to compute $\avg{\Delta U}_N$ using the probability in \eqnref{eq:probcc_gen}, we now choose the following engine Hamiltonian:
\begin{align}
  H_E(t) = 2\Omega(t)\, S_z + 2 \Delta S_x\,, \label{eq:he}
\end{align}
with $S_z \equiv (a^\dagger a - b^\dagger b)/2$. In the Otto cycle, both the compression and expansion strokes are done without coupling to the heat baths, and, hence, they are described by unitary dynamics governed by the engine Hamiltonian (\ref{eq:he}). For quasistatic changes of $\Omega(t)$, the propagator can be written as
\begin{align}
U_R(t,t_0) \approx  \sum_{m=-N/2}^{N/2} \ket{m,\theta_t}_E {_E}\bra{m,\theta_{t_0}}\, e^{-i m \phi(t,t_0)},\label{eq:adbprop}
\end{align}
with $\phi(t,t_0) = \int_{t_0}^{t} d t^{\prime}\, 2 E_{t^{\prime}}$. Here, we denote by $|m, \theta_t\rangle_E$  the eigenstate of  the instantaneous engine Hamiltonian $H_E(t)$ with eigenvalue  $2E_tm$, where  $E_t \equiv \sqrt{\Omega(t)^2 + \Delta^2}$. Furthermore,  $\theta_t$ is defined by $\tan{\theta_t} = - \Omega(t)/\Delta$. The initial time is $t_0 = 0$ for the isentropic compression and $t_0 = T/2$ for the isentropic expansion strokes. With this adiabatic propagator, we obtain the autocorrelation function of the operator $V_R$ (see \cite{Supplement} for details), when $t_0 \leq \{t,t^{\prime}\}\leq t_0+T/2$, in \eqnref{eq:probcc_gen} as \cite{note:autocorrfunc}
\begin{align}
&\langle V_R^{(I)}(t^{\prime})\, V_R^{(I)}(t) \rangle_{\rho_{t_0}^R} = 4 \cos{\theta_{t}}\, \cos{\theta_{t^{\prime}}}\, \avg{m^2}_{t_0} \label{eq:twotimecorr_indisting}\\ 
&\quad +\sin{\theta_{t}}\, \sin{\theta_{t^{\prime}}} \sum_{\sigma=\pm} e^{-i \sigma \phi(t^{\prime},t)} \left[\frac{N}{2}\left(\frac{N}{2}+1 \right)-F_{\sigma}(N,\beta_{t_0}E_{t_0}) \right],\nonumber
\end{align}
with $\beta_0 \equiv \beta_c$, $\beta_{T/2} \equiv \beta_h$, and
$F_{\pm}(N,\beta_{t_0}E_{t_0}) \equiv \langle m^2\rangle_{t_0}  \pm\langle m\rangle_{t_0}$, where the expectation values are defined with respect to the thermal state of $H_E(t_0)$ at $\beta_{t_0}$. On the other hand, if $t$ and $t^{\prime}$ are separated by the thermalization process at $t=T/2$, for instance $t < T/2$ and $t^{\prime} > T/2$, the autocorrelation function takes the factorized form $\langle V_R^{(I)}(t^{\prime})\, V_R^{(I)}(t) \rangle_{\rho^R_0} = \langle V_R^{(I)}(t^{\prime}) \rangle_{\rho^R_{T/2}} \langle V_R^{(I)}(t) \rangle_{\rho^R_0}$ with
\begin{align}
\langle V_R^{(I)}(t) \rangle_{\rho^R_{t_0}} = 2 \cos{\theta_{t}}\, \langle m\rangle_{t_0}\,. \label{eq:singopavg_indist}
\end{align}

Next, we examine the case in which all the atoms are distinguishable. In this case, the coupling Hamiltonian (\ref{eq:hc_sx}) reduces to $H_C(t) = g_C(t)\, 2 \sum_{j=1}^N (\sigma_{j,\, x}/2) \otimes V_S$ and the engine Hamiltonian (\ref{eq:he}) to $H_E(t) = 2\Omega(t) \sum_{j=1}^N (\sigma_{j,\, z}/2) + 2\Delta \sum_{j=1}^N (\sigma_{j,\, x}/2)$, where $\sigma_{j,\, x}$ and $\sigma_{j,\, z}$ are the Pauli matrices of the $j$th atom. Assuming quasistatic changes of $\Omega(t)$ like in the indistinguishable case, we find the following autocorrelation function:
\begin{align}
\langle V_R^{(I)}(t^{\prime})\, V_R^{(I)}(t) \rangle_{\rho_{t_0}^R} =&  \cos{\theta_{t}} \cos{\theta_{t^{\prime}}} \left[N+N(N-1)\tanh^2{(\beta_{t_0}E_{t_0})} \right] \nonumber\\ 
  &+ \frac{N}{2} \frac{\sin{\theta_{t}} \sin{\theta_{t^{\prime}}}}{\cosh{(\beta_{t_0}E_{t_0})}} \sum_{\sigma=\pm} e^{\sigma\left [ i\phi(t^{\prime},t)-\beta_{t_0}E_{t_0} \right]}\,,\label{eq:twotimecorr_disting}
\end{align}
and the average
\begin{align}
\langle V_R^{(I)}(t) \rangle_{\rho_{t_0}^R} = -N \cos{\theta_{t}}\, \tanh{(\beta_{t_0} E_{t_0})}\,, \label{eq:singopavg_dist}
\end{align}
allowing the calculation of the probability in \eqnref{eq:probcc_gen} for the distinguishable case. We emphasize that in the distinguishable case, while taking the trace over the engine states, all the possible $2^N$ configurations of the atomic pseudospins have to be considered, while in the indistinguishable case the trace is taken over only $N+1$ symmetrized eigenstates of $S_z$. Thus, the collective nature of the latter set of states gives rise to an enhanced coupling with the external system and results in the enhancement of $\Delta U$ that we demonstrate next.

\begin{figure}[t]
	\centering
	\includegraphics[width= \columnwidth]{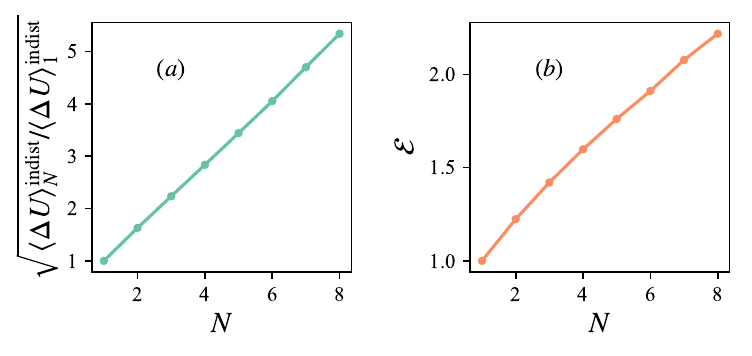}
	\caption{
		{Quantum performance of multiple identical engines with continuous nonperturbative coupling.} (a) $\avg{\Delta U}_N^{\rm indist}$ done by $N$ indistinguishable engines outcoupled via a continuous nonperturbative coupling. A quadratic scaling is shown  for small $N$.  (b) Scaling of the enhancement $\mathcal{E}$.  Engine parameters are $v = 0.1 \Omega(0)^2$, $T = 20/\Omega(0)$, $\Delta = 0$, and $\Omega(0) = 1$. The coupling $g_C (t)$, to the external HO system of frequency $\omega = 2\pi \times 0.05/T$, is chosen with $g = 0.5$, $\delta_t = 0.9$, $\alpha = 2142/T$, $t_{\rm on} = (1-\delta_t)T/4$, and $t_{\rm off} = t_{\rm on}+\delta_t T/2$.}
	\label{fig:bose_identical_cc_nonpert}
\end{figure}

To clearly evidence that Bose statistics leads to a quantum advantage, we specialize the coupling protocol to the impulse form $g_C(t) = g \delta(t-t_1)$, with $0 < t_1 < T/2$. In this case, the expressions for the probability (\ref{eq:probcc_gen}) are simplified greatly, and the average internal energy change equals
\begin{align}
\avg{\Delta U}_N \simeq g^2 \left\langle [V_R^{(I)}(t_1)]^2 \right\rangle_{\rho^R_0}\,\, \sum_{i \neq 0} \epsilon_i^S\, \left|{_S}\langle i| V_S^{(I)}(t_1) |0\rangle_S \right|^2\label{eq:wndelta}.
\end{align}
Thus, for the impulse form of the coupling, $\avg{\Delta U}_N$ for the indistinguishable and distinguishable cases differ by the value of the variance $\langle [V_R^{(I)}(t_1)]^2 \rangle_{\rho^R_0}$ that can be evaluated from Eqs. \eqref{eq:twotimecorr_indisting} and \eqref{eq:twotimecorr_disting}.
Remarkably, we find that, in the indistinguishable case, this variance is larger than or equal to that of the distinguishable case for any values of the parameters $N \geq 1$, $\beta_c E_0$, and $\theta_{t_1}$ (see \cite{Supplement}). This fact guarantees that $\avg{\Delta U}_N^{\text{indist}}$ due to $N$ indistinguishable bosonic engines is always larger than $\avg{\Delta U}_N^{\text{dist}}$ arising from the same number of distinguishable engines. The resulting enhancement can be quantified by the ratio $\mathcal{E}=\avg{\Delta U}_N^{\text{indist}}/\avg{\Delta U}_N^{\text{dist}}$.

In \figref{fig:bose_identical_delta} (a), we compare our analytical result for the enhancement $\mathcal{E}$, for different values of $\theta_{t_1}$ and $N$, with numerical simulations for a specific choice of the system as a HO with frequency $\omega$, i.e., $H_S=\omega c^\dagger c$ and $V_S = c^\dagger + c$ in the coupling Hamiltonian (\ref{eq:hc_sx}). For the engine Hamiltonian $H_E(t)$, we consider linear sweeps of $\Omega(t)$ as $\Omega(t)=\Omega(0)+vt$ and $\Omega(t)=\Omega(0)+v(T-t)$ for the isentropic compression and expansion strokes, respectively. Hereafter, let us focus on the situation with $\Delta = 0$ (i.e., $\theta_{t}= - \pi/2$), where the difference between the indistinguishable and distinguishable cases is most prominent \cite{note:adiabaticity}. When $\Delta = 0$, using \eqnref{eq:twotimecorr_disting} we see that $\avg{\Delta U}_N \propto N$ for the distinguishable case, while in the indistinguishable case using \eqnref{eq:twotimecorr_indisting} the dependence on $N,\beta_c E_0$ is more involved (see \cite{Supplement}). In general, although $\langle [V_R^{(I)}(t_1)]^2 \rangle_{\rho^R_0}$ contains both terms proportional to $N^2$ and $N$, we find that $\avg{\Delta U}_N$ shows $N^2$ scaling for moderate values of $N$ with $N \beta_c \Omega(0) \alt 1$ \cite{note:dicke}. We see this behavior in \figref{fig:bose_identical_delta} (b), where we plot $\sqrt{\avg{\Delta U}_N/\avg{\Delta U}_1}$ to bring out the quadratic scaling. We also note that, for sufficiently large $N$, the $N^2$ scaling of $\langle [V_R^{(I)}(t_1)]^2 \rangle_{\rho^R_0}$ for the indistinguishable case turns into a linear scaling with an enhanced slope of $\coth{\beta_c\Omega(0)}>1$, while it is unity in the distinguishable case. 

Considering general coupling protocols $g_C(t)$, when $\Delta = 0$, we find that the autocorrelation $\langle V_R^{(I)}(t^{\prime})\, V_R^{(I)}(t) \rangle_{\rho^R_0}$ is factorized, and hence, vanishes when $t$ and $t^{\prime}$ are separated by a thermalization step with the hot bath.
This allows us to simplify \eqref{eq:probcc_gen} and write the probability of excitation of the driven system in the indistinguishable case as
\begin{align}
p_i^{\mr{indist}} &\simeq \sum_{\substack{t_0 = {0,T/2} \\ \sigma=\pm}} \vert c^{\sigma}_i(t_0) \vert^2 \left[\frac{N(N+2)}{4}-F_{\sigma}(N,\beta_{t_0}E_{t_0}) \right] \label{eq:wNdel0_indist}
\end{align}
and in the distinguishable case as
\begin{align}
p_i^{\mr{dist}} &\simeq \sum_{\substack{t_0 = {0,T/2} \\ \sigma=\pm}}\vert c^{\sigma}_i(t_0) \vert^2 \left[\frac{N}{2}(1+\sigma \tanh{(\beta_{t_0}E_{t_0})}) \right] \label{eq:wNdel0_dist},
\end{align}
where the positive, coupling-protocol-dependent terms are determined by the amplitudes $c^{\pm}_i(t_0) = \int_{t_0}^{t_0+T/2} dt\, g_C(t)\,\, {_S}\sandwich{i}{V_S^{(I)}(t)}{0}_S\, e^{\pm i \phi(t,t_0)}$. 
Comparing the terms in brackets in Eqs.~(\ref{eq:wNdel0_indist}) and (\ref{eq:wNdel0_dist}), we have that $p_i^{\mr{indist}}\geq p_i^{\mr{dist}}$.
Thus, for $\Delta = 0$, our central result of the enhancement of internal energy change for indistinguishable bosonic engines still holds for arbitrary coupling protocols and external system Hamiltonians.

In order to widen the scope of our results, we also consider engine strokes with $\Delta \neq 0$ as described in more detail in Ref.~\cite{Supplement}. There, we find that the enhancement persists for small values of $N$ independently of the form of the coupling and the external system Hamiltonian. The fact that the enhancement is guaranteed for small $N$ most likely will be of particular relevance to experiments in the near future that will presumably have access to small $N$.
Furthermore, we demonstrate for a given choice of the Hamiltonian how to identify generic parameter regimes and coupling protocols that lead to enhancement. We also extend our results to nonperturbative continuous coupling $g_C (t)$ [see \cite{Supplement} for the exact functional form of $g_C(t)$] using numerical simulations for a HO external system. From \figref{fig:bose_identical_cc_nonpert}, we see that, for small values of $N$ and moderate values of the coupling strength, there is enhancement and $\avg{\Delta U}_N^{\rm indist}$ scales as $N^2$. The fact that we are able to find enhancement for a generic set of parameters as in \figref{fig:bose_identical_cc_nonpert} without fine-tuning suggests the general applicability of our result. Finally, we note that the performance in the case of engines made of $N$ identical noninteracting fermionic two-level atoms, considered in Ref.~\cite{Supplement}, is generally diminished for even $N$ and converges to that by a single engine for odd $N$ in the limit of small $\beta_{\rm COM}^{-1}$ due to the Pauli blocking effect.

In conclusion, we have demonstrated that the statistical indistinguishability of work resources can be exploited to gain a genuine quantum enhancement in quantum thermodynamics. While we identify this enhancement in terms of the internal energy change of an external system coupled to the engines, the question arises as to how much of this change is attributable to the actual action of the engines and how much results from the time dependence of the part of the Hamiltonian describing the interaction between the system and the engines. Our preliminary analysis indicates that an accordingly corrected work contribution of engines also displays enhancement in the parameter regimes considered here \cite{Supplement}. The predicted enhancement of the energy output from multiple indistinguishable heat engines to a generic external system is readily testable with current or near-future experimental realizations of quantum heat engines, e.g., in nitrogen-vacancy centers \cite{Klatzow19}, trapped ions \cite{vonlindenfels}, and ultracold gases. While we have considered bosonic and fermionic statistics \cite{Supplement}, exotic fractional statistics \cite{khare05,chen18} may lead to further interesting results.

\bigskip
\begin{acknowledgments} 
We thank Yanming Che, I\~nigo L. Egusquiza and Kosuke Ito for helpful discussions and comments. 
G.~W. is supported by the Zhejiang Provincial Natural Science Foundation Key Project (Grant No. LZ19A050001), by National Natural Science Foundation of China (Grants No. 11975199 and No. 11674283), by the Fundamental Research Funds for the Central Universities (2017QNA3005 and 2018QNA3004), and by the Zhejiang University 100 Plan.
B.~P.~V. is supported by the Research Initiation Grant, Excellence-in-Research Fellowship of IIT Gandhinagar, and a Department of Science \& Technology-Science and Engineering Research Board (India) Start-up Research Grant No. SRG/2019/001585.
M.~J.~H. is supported by the startup fund by Duke Kunshan University and the European Research Council Synergy grant BioQ.
G.~W. and B.~P.~V. led the project and contributed equally.
\end{acknowledgments}

\clearpage
\onecolumngrid

\begin{center}
	
	\newcommand{\beginsupplement}{%
		\setcounter{table}{0}
		\renewcommand{\thetable}{S\arabic{table}}%
		\setcounter{figure}{0}
		\renewcommand{\thefigure}{S\arabic{figure}}%
	}
	
        \textbf{\large --- Supplemental Material --- \\Quantum Statistical Enhancement of the Collective Performance of Multiple Bosonic Engines}
\end{center}
\newcommand{\beginsupplement}{%
	\setcounter{table}{0}
	\renewcommand{\thetable}{S\arabic{table}}%
	\setcounter{figure}{0}
	\renewcommand{\thefigure}{S\arabic{figure}}%
}

\setcounter{equation}{0}
\setcounter{figure}{0}
\setcounter{table}{0}
\setcounter{page}{1}
\makeatletter
\renewcommand{\theequation}{S\arabic{equation}}
\renewcommand{\thefigure}{S\arabic{figure}}
\renewcommand{\bibnumfmt}[1]{[S#1]}
\renewcommand{\citenumfont}[1]{S#1}
\vspace{0.8 in}

\newcommand{\D}{\Delta}
\newcommand{\tD}{\tilde{\Delta}}
\newcommand{\K}{K_{PP}}
\newcommand{\bn}{\bar{n}_P}
\newcommand{\G}{\Gamma}
\newcommand{\LH}{\underset{L}{H}}
\newcommand{\HL}{\underset{H}{L}}

\newcommand{\ketind}[2]{\ket{#1,{\theta}_{#2}}_E}
\newcommand{\braind}[2]{{}_E\bra{#1,{\theta}_{#2}}}
\newcommand{\ketdis}[3]{\ket{#1,{\theta}_{#2}}_{#3}}
\newcommand{\bradis}[3]{{}_{#3}\bra{#1,{\theta}_{#2}}}
\newcommand{\sxs}[1]{\sigma_{#1,\, x}}
\newcommand{\sys}[1]{\sigma_{#1,\, y}}
\newcommand{\szs}[1]{\sigma_{#1,\, z}}
\newcommand{\sps}[1]{\sigma_{#1,\, +}}
\newcommand{\sms}[1]{\sigma_{#1,\, -}}
\newcommand{\md}{m^{\prime}}
\newcommand{\tp}{t^{\prime}}
\newcommand{\pvmod}[1]{{\color{blue}#1}}


\vspace{-2 cm}

\subsection{Performance of multiple engines: derivation of Eqs.~(8)--(11)}
Let us begin by considering the indistinguishable case. The adiabatic propagator (propagator under quasistatic changes of external parameters) given by Eq.~(7) in the main paper has been written in terms of the instantaneous energy eigenstates, which can be expressed as
\begin{align}
\ketind{m}{t} = e^{-i (\theta_t+\frac{\pi}{2})S_y} \ket{m}_z\,, \label{eq:ketcolldef}
\end{align} 
with $S_z \ket{m}_z = m \ket{m}_z$. Note that, in what follows, the variable $m$ takes values between $-N/2 \leq m \leq N/2$. For ease of presentation, we will not show the range of $m$ in the summations that appear below. Using the above form for the eigenstates, the operator $V_R^{(I)}(t)$ in the coupling Hamiltonian can be written as
\begin{align}
V_R^{(I)}(t) &= 2 U_R^{\dagger}(t,t_0)\, S_x\, U_R(t,t_0)\nonumber\\
&= 2 \sum_{m,m^{\prime}} e^{i(\md-m)\,\phi(t,t_0)}\,_z\sandwich{\md}{e^{i(\theta_t+\frac{\pi}{2})S_y}\, S_x\, e^{-i(\theta_t+\frac{\pi}{2})S_y}}{m}_z\, \ketind{\md}{t_0} \braind{m}{t_0} \nonumber \\
&=2 \sum_{m,m^{\prime}} e^{i(\md-m)\,\phi(t,t_0)}\,_z\sandwich{\md}{(-\sin{\theta_t}\, S_x + \cos{\theta_t}\, S_z)}{m}_z\, \ketind{\md}{t_0} \braind{m}{t_0}   \nonumber \\
&= 2 \cos{\theta_t}\, S_z(t_0) - \sin{\theta_t} \left[S_+(t_0)\, e^{i\phi(t,t_0)} + S_-(t_0)\, e^{-i\phi(t,t_0)} \right]\,, \label{eq:Vrindistsimp}
\end{align}
with $S_{\alpha}(t_0) =e^{-i(\theta_{t_0}+\frac{\pi}{2})S_y}\, S_{\alpha}\, e^{i(\theta_{t_0}+\frac{\pi}{2})S_y}$, $S_{\pm}\equiv (S_x \pm i S_y)$, and $\phi(t,t_0)$ being the dynamical phase acquired during the adiabatic evolution from time $t_0$ to $t$ given by
\begin{align}
  \phi(t,t_0) = \int_{t_0}^t dt^{\prime}\, 2 E_{t^{\prime}}\,.\nonumber
\end{align}
We can now use the rotated collective spin operators to express the initial Hamiltonian of the engine as $H_E(t_0) = 2 E_{t_0} S_z(t_0)$ and the density matrix of the engines at $t_0$ is $\rho^E_{t_0} = e^{-2\beta_{t_0} E_{t_0}S_z(t_0)}/Z_{\beta_{t_0}}$, allowing us to calculate the expectation value of the coupling operator as
\begin{align}
\avg{V_R^{(I)}(t)}_{\rho^R_{t_0}} &= \Tr \left[ V_R^{(I)} (t) \rho^E_{t_0}\right] \nonumber\\
&= 2 \cos{\theta_t}\,\frac{\sum_m m e^{-2m \beta_{t_0} E_{t_0}}}{Z_{\beta_{t_0}}}  = 2 \cos{\theta_t}\,\avg{m}_{t_0}\,, \label{eq:deriveq10}
\end{align}
which proves Eq.~(9) in the main paper. Note that the partition function in the indistinguishable case is given by $Z_{\beta_{t_0}} = e^{N\beta_{t_0}E_{t_0}}(1-e^{2(N+1)\beta_{t_0}E_{t_0}})/(1-e^{2\beta_{t_0}E_{t_0}})$. In order to compute the correlation function given in Eq.~(8), consider first the operator product
\begin{align}
V_R^{I}(\tp)\, V_R^{I}(t) =& 4 \cos{\theta_t}\, \cos{\theta_{\tp}}\, S_z^2(t_0) + \sin{\theta_t}\, \sin{\theta_{\tp}} \left[S_+(t_0)\, S_-(t_0)\, e^{i \phi(\tp,t)} + S_-(t_0)\, S_+(t_0)\, e^{-i \phi(\tp,t)} \right] \nonumber\\
&+ \sin{\theta_t}\, \sin{\theta_{\tp}} \left[ S_+^2(t_0)\, e^{i \phi(t,t_0) + i \phi(\tp,t_0)} + \mathrm{h.c.} \right] - 2 \cos{\theta_{\tp}}\, \sin{\theta_t}\, S_z(t_0)\left[S_+(t_0)\, e^{i\phi(t,t_0)} + \mathrm{h.c.}\right]\nonumber\\
& - 2 \cos{\theta_{t}}\, \sin{\theta_{\tp}} \left[S_+(t_0)\, e^{i\phi(t,t_0)} + \mathrm{h.c.}\right]S_z(t_0)\,,\label{eq:operatorcorrind}
\end{align}
with ``h.c.'' denoting Hermitian conjugate. Since the density matrix at $t_0$ is diagonal in the instantaneous basis $\ket{m,\theta_{t_0}}$, only the first line of the above equation contributes to the trace. Taking into account the following relations, $S_{+}(t_0)\, S_{-}(t_0) = \frac{N}{2}(\frac{N}{2}+1)-[S_z^2(t_0)-S_z(t_0)]$ and $S_{-}(t_0)\, S_{+}(t_0) = \frac{N}{2}(\frac{N}{2}+1)-[S_z^2(t_0)+S_z(t_0)]$, and defining
\begin{align}
\avg{m^2}_{t_0} \equiv \frac{\sum_m m^2 e^{-2m \beta_{t_0} E_{t_0}}}{Z_{\beta_{t_0}}}\,, \label{eq:secondmoment}
\end{align}
we obtain the expression (8) of the main paper for $\avg{V_R^{I}(\tp)\, V_R^{I}(t)}_{\rho^{R}_{t_0}}$.

Let us now consider the distinguishable case. As mentioned in the main paper, the time-dependent engine Hamiltonian in this case is given by
\begin{align}
H_E(t) = \sum_{j=1}^{N} H_{E,j} (t) = \sum_{j=1}^{N} \left[\Omega(t) \szs{j}+\Delta \sxs{j}\right]\,, 
\end{align}
and the resulting propagator in the quasistatic limit is simply given by a direct product over individual engine propagators:
\begin{align}
U_R(t,t_0) \approx \bigotimes_{j=1}^N \sum_{m=-1/2}^{1/2} \ketdis{m}{t}{j} \bradis{m}{t_0}{j}\, e^{-i m \phi(t,\, t_0)}\,. \label{eq:distprop}
\end{align}
Here, $\ketdis{m}{t}{j}$ is the instantaneous eigenstate of the $j$th individual engine satisfying $H_{Ej}(t) \ketdis{m}{t}{j} = 2 E_t \ketdis{m}{t}{j}$, which is given by
\begin{align}
\ketdis{m}{t}{j} = e^{-i(\theta_t+\pi/2)\sys{j}/2}\ket{m}_{jz}\,,
\end{align}
with $\szs{j} \ket{m}_{jz} = 2m \ket{m}_{jz}$. We can now repeat the calculation in a manner similar to the indistinguishable case. To this end, we first note that the coupling operator in the interaction picture is given by
\begin{align}
V_R^{(I)}(t) = \sum_{j=1}^N \left\{\cos{\theta_t}\, \szs{j}(t_0) - \sin{\theta_t} \left[\sps{j}(t_0)\, e^{i\phi(t,t_0)} + \sms{j}(t_0)\, e^{-i \phi(t,t_0)}\right]\right\}\,, \label{eq:VRIdist}
\end{align}
with $\sigma_{j,\alpha}(t_0) = e^{-i(\theta_{t_0}+\pi/2)\sys{j}/2} \sigma_{j,\alpha} e^{i(\theta_{t_0}+\pi/2)\sys{j}/2}$ and $\sigma_{\pm}\equiv (\sigma_x \pm i \sigma_y)/2$. Using the above expression and the fact that the density matrix for the distinguishable case is given by
\begin{align}
\rho^E_{t_0} = \frac{\prod_{j=1}^{N} e^{-\beta_{t_0}E_{t_0}\szs{j}(t_0)}}{\cosh^N(\beta_{t_0}E_{t_0})}\,,
\end{align}
we obtain the average of the coupling operator (noting that only the first term from \eqnref{eq:VRIdist} contributes as the density operator is diagonal in the $\ketdis{m}{t}{j}$ basis)
\begin{align}
\avg{V_R^{(I)}(t)}_{\rho_{t_0}^R} = \frac{\sum_{j=1}^{N} \Tr [\cos{\theta_t}\, \szs{j}(t_0)\, e^{-\beta_{t_0}E_{t_0}\szs{j}(t_0)}]}{\cosh^N{(\beta_{t_0}E_{t_0})}}\,.
\end{align}
This leads to Eq.~(11) of the main paper since each term in the sum is the same and the trace over everything but the $j$th engine cancels with the partition function in the denominator to give $\Tr{[\szs{j}(t_0)\, \rho^E_{t_0}]} = -\tanh{(\beta_{t_0} E_{t_0})}$. As before, we can again write down the two-time correlation operator for the coupling and many of the terms vanish while taking the average due to the diagonal nature of the density matrix. Finally, we get for the correlation function
\begin{align}
\avg{V_R^{(I)}(\tp)\, V_R^{(I)}(t)}_{\rho_{t_0}^R} =& \sum_{j=1}^N \cos{\theta_t}\, \cos{\theta_{\tp}}\, \Tr[ \szs{j}(t_0)\, \szs{j}(t_0)\, \rho^E_{t_0}] + \sum_{j \neq k} \cos{\theta_t}\, \cos{\theta_{\tp}}\, \Tr[ \szs{j}(t_0)\, \szs{k}(t_0)\, \rho^E_{t_0}] \nonumber \\
&+ \sin{\theta_t}\, \sin{\theta_{\tp}} \sum_{j=1}^N \left[ \Tr [\sps{j}(t_0)\, \sms{j}(t_0)\, \rho^E_{t_0}]\, e^{i\phi(\tp,t)} +\Tr [\sms{j}(t_0)\, \sps{j}(t_0)\, \rho^E_{t_0}]\, e^{-i\phi(\tp,t)} \right]. \label{eq:corrDistexpression}
\end{align}
Notice that in the first line of the above equation, the first sum has $N$ equal terms and the second sum has $N(N-1)$ equal terms, and the second line of the equation has $N$ equal terms. Combining this with the fact that $\Tr{[\sps{j}(t_0)\, \sms{j}(t_0)\, \rho^E_{t_0}]} = e^{-\beta_{t_0} E_{t_0}}/[2 \cosh{(\beta_{t_0} E_{t_0})}]$ and $\Tr{[\sms{j}(t_0)\, \sps{j}(t_0)\, \rho^E_{t_0}]} = e^{\beta_{t_0} E_{t_0}}/[2 \cosh{(\beta_{t_0} E_{t_0})}]$ leads to Eq.~(10) of the main paper.

\subsection{Impulse-type coupling between the engines and the system}
 The average internal energy change with impulse-type coupling at time $t_1$ (assuming $0 < t_1 < T/2$) is discussed in the main paper, see Eq.~(12). This discussion carries over to the case $T/2 < t_1 < T$ up to the fact that the variance has to be taken with respect to the work resource state $\rho^R_{T/2}$ and is given by
\begin{align}
\avg{\Delta U}_N \simeq g^2 \left\langle [V_R^{(I)}(t_1)]^2 \right\rangle_{\rho^R_0}\, \sum_{i \neq 0} \epsilon_i^S\, \left|{_S}\langle i| V_S^{(I)}(t_1) |0\rangle_S \right|^2.\label{eq:wndelta_supp}
\end{align}
An important simplification arising from the impulse coupling becomes evident in the above equation:
the average internal energy change is a product of a term that depends only on the engine or work-resource operators and a term that contains the system operators alone. The engine-dependent quantity is the second moment of the coupling  $\langle [V_R^{(I)}(t_1)]^2 \rangle_{\rho^R_0}$, and it can be obtained using the general expression for the two-time correlators $\langle V_R^{(I)}(t) V_R^{(I)}(t^{\prime}) \rangle_{\rho^R_0}$ provided in Eqs.~(8) and (10) of the main paper as
\begin{align}
\left\langle [V_R^{(I)}(t_1)]^2 \right\rangle_{\rho^R_0} =  \left[ \frac{1}{2} N(N+2) - 2 \avg{m^2}_{0} \right] \sin^2{\theta_{t_1}} + 4 \avg{m^2}_{0} \cos^2{\theta_{t_1}}\,\label{eq:vrsq_indist2}
\end{align}
for the indistinguishable case and as
\begin{align}
\left\langle [V_R^{(I)}(t_1)]^2 \right\rangle_{\rho^R_0} = N \sin^2{\theta_{t_1}} +\left[N + N(N-1)  \tanh^2{(\beta_c E_0)} \right] \cos^2{\theta_{t_1}} \label{eq:vrsq_dist2}
\end{align}
for the distinguishable case. Let us now introduce the function $f(N,\beta_{t_0} E_0) \equiv \langle m^2\rangle_{t_0}$ to explicitly account for the functional dependence on the expectation value of $m^2$ (recall that $t_0 = 0,T/2$ and $\beta_0 = \beta_c$ and $\beta_{T/2}=\beta_h$): 
\begin{align}
  f(N, \beta_{t_0}E_{t_0}) &\equiv\frac{\displaystyle \sum_{m=-N/2}^{N/2} m^2 e^{-\beta_{t_0} 2 E_{t_0} m}}{Z_{\beta_{t_0}}}\nonumber\\
  &=\frac{\langle H_E^2(0)\rangle_{\rho_0^E}}{4E_{t_0}^2}\nonumber\\
  &= \frac{ N^2\, \sinh{[(N+3) \beta_{t_0} E_{t_0}]} + (N+2)^2\, \sinh{[(N-1) \beta_{t_0} E_{t_0}]}-2\left(N^2+2N-2\right)\sinh{[(N+1) \beta_{t_0} E_{t_0}]}} {16 \sinh{[(N+1) \beta_{t_0} E_{t_0}]} \sinh^2{(\beta_{t_0} E_{t_0})}}\,.\label{eq:f}
\end{align}	
Now, using the inequalities
\begin{align}
f(N,x) &\leq N^2/4 \label{eq:cond1a},\\
4 f(N,x) &\geq \left[N+N(N-1)\tanh^2{(x)} \right], \label{eq:cond1_supp}
\end{align}
we compare the coefficients of the $\sin^2{\theta_{t_1}}$ and $\cos^2{\theta_{t_1}}$ terms in Eqs. \eqref{eq:vrsq_indist2} and \eqref{eq:vrsq_dist2}. It follows that the second moment $\langle [V_R^{(I)}(t_1)]^2 \rangle$, and thus the average internal energy change as well, are larger for the indistinguishable case, as we stated in the main text.

Let us now consider $\Delta = 0$ (i.e., $\cos{\theta_{t_1}} = 0$) and examine the scaling of the average internal energy change with $N$. To this end, note from \eqnref{eq:wndelta_supp} that it is sufficient to examine the  scaling of the second moment with $N$. Nonetheless,  to make the  comparison with numerical solutions as in Fig.~2(b) of the main paper, we note that when the external system is chosen as a harmonic oscillator with frequency $\omega$,  one finds ${_S}\wich{i}{V_S^{(I)}(t)}{0}_S = e^{i\omega t} \delta_{i,1}$ and the average internal energy change (\ref{eq:wndelta_supp}) becomes  
\begin{align}
\avg{\Delta U}_N \simeq \omega g^2 \left\langle [V_R^{(I)}(t_1)]^2 \right\rangle_{\rho^R_0}.
\end{align}
Turning our attention to the second moment for the distinguishable case, we first see that $\langle [V_R^{(I)}(t_1)]^2 \rangle_{\rho^R_0} = N$. To examine the behavior of the variance in  \eqnref{eq:vrsq_indist2} for the indistinguishable case, consider first Eq.~(\ref{eq:f}) in the limit of $N\rightarrow \infty$ 
\begin{align}
f(N,\, \beta_cE_0) \rightarrow  \frac{N^2}{4} + \frac{N}{2} \left[1-\coth{(\beta_c E_0)}\right] + \frac{1}{2} \left[\coth{(\beta_c E_0)} - 1\right] \coth{(\beta_c E_0)}\,,
\end{align}
that yields
\begin{align}
\langle [V_R^{(I)}(t_1)]^2 \rangle_{\rho^R_0} \rightarrow \bigl\{ \coth{(\beta_c E_0)}\, N - \left[\coth{(\beta_cE_0)} -1 \right] \coth{(\beta_cE_0)} \bigr\}, \label{eq:vrsq_indist_ninf}
\end{align}
and scales linearly with $N$ with a slope $\coth(\beta_c E_0)\geq 1$, as stated in the main paper. This term is thus larger in the indistinguishable case than in the distinguishable counterpart and is responsible of the quantum statistical enhancement of the average internal energy change. In the other extreme with $N=1$, since $f(N=1,\, \beta_cE_0) = 1/4$,  the second moment in the indistinguishable and distinguishable cases are both equal to $1$, as expected. For $N \beta_c\Omega(0) \sim 1$ or less we have, $\langle [V_R^{(I)}(t_1)]^2 \rangle_{\rho^R_0} \simeq f_1(\beta_c\Omega(0))\, N^2 + f_2(\beta_c\Omega(0))\, N$, with $f_1(x) \equiv x (x \coth{x}-1) \cosh{x}/\sinh^3{x}$ and $f_2(x) \equiv 1 - [(x \coth{x}-1)/\sinh^2{x}]$. Therefore, $\avg{\Delta U}_N$ shows $N^2$ scaling for moderate values of $N$ with $N \beta_c\Omega(0) \alt 1$ for the indistinguishable case and is larger than the distinguishable case which scales linearly with $N$.

\subsection{General coupling between the engines and the system}
For a general continuous coupling, the probability of exciting the $i$th eigenstate of the system reads
\begin{align}
p_{i} \simeq \int_0^T dt \int_0^T dt^{\prime}\, g_C(t)\, g_C(t^\prime)\,\,_S\wich{i}{V_S^{(I)}(t)}{0}_S \,_S\wich{0}{V_S^{(I)}(t^{\prime})}{i}_S\,\, \langle V_R^{(I)}(t^{\prime}) V_R^{(I)}(t) \rangle_{\rho_0^R}\,. \label{eq:probcc_gen_supp}
\end{align} 
Using the two-time correlators and one-time averages stated in the main paper (see Eqs.~(8)--(11) there), the excitation probability for the case of many indistinguishable and distinguishable engines can be respectively written as
\begin{align}
p_i^{\mr{indist}} =& \sum_{t_0=0,T/2} 4 \vert d_i(t_0) \vert^2 \avg{m^2}_{t_0} + \vert \tilde{c}^{+}_i(t_0) \vert^2 \left[\frac{N}{2}\left(\frac{N}{2}+1 \right)-F_{+}(N,\beta_{t_0}E_{t_0}) \right] + \vert \tilde{c}^{-}_i(t_0) \vert^2  \left[\frac{N}{2}\left(\frac{N}{2}+1 \right)-F_{-}(N,\beta_{t_0}E_{t_0})\right] \nonumber \\
&+ 8 \Re [d_i(0)d_i^*(T/2)]\, \avg{m}_{0}\avg{m}_{T/2}\,, \label{eq:probgen_indist}\\
p_i^{\mr{dist}} =& \sum_{t_0=0,T/2} \vert d_i(t_0) \vert^2 \left[N+N(N-1)\tanh^2{(\beta_{t_0} E_{t_0})} \right] + \vert \tilde{c}^{+}_i(t_0) \vert^2 \left[\frac{N}{2}(1+\tanh{(\beta_{t_0}E_{t_0})}) \right] + \vert \tilde{c}^{-}_i(t_0) \vert^2  \left[\frac{N}{2}(1-\tanh{(\beta_{t_0}E_{t_0})})\right] \nonumber \\
&+  2 \Re [d_i(0)d_i^*(T/2)]N^2\tanh{(\beta_c E_0)} \tanh{(\beta_h E_{T/2})}\,. \label{eq:probgen_dist}
\end{align}
Here, the coupling-protocol- and engine-state-dependent amplitudes  are given by 
\begin{align}
\tilde{c}^{\pm}_i(t_0) &= \int_{t_0}^{t_0+T/2} dt\, g_C(t) \sin{\theta_t}\, {_S}\sandwich{i}{V_S^{I}(t)}{0}_S\, e^{\pm i \phi(t,t_0)},\\
d_i(t_0) &= -\int_{t_0}^{t_0+T/2} dt\, g_C(t) \cos{\theta_t}\, {_S}\sandwich{i}{V_S^{I}(t)}{0}_S,
 \end{align}
and generalize the expressions presented in the main paper. Similar to the function $f(N,x)$ defined in \eqnref{eq:f}, we now define $h(N,\beta_{t_0}E_{t_0}) \equiv \avg{m}_{t_0}$ to account for the parameter dependence of the average $\avg{m}$. This gives
	\begin{align}
	h(N,\beta_{t_0}E_{t_0}) &\equiv \sum_{m=-N/2}^{N/2} m\, \frac{e^{-\beta_{t_0}2E_{t_0}m}}{Z_{\beta_{t_0}}} \nonumber\\
	&=\frac{\langle H_E(0)\rangle_{\rho_0^E}}{2E_{t_0}}\nonumber\\
	&=\frac{1}{4} \frac{(N+2) \sinh{(N\beta_{t_0}E_{t_0})} - N \sinh{[(N+2)\beta_{t_0}E_{t_0}]}}{\sinh{(\beta_{t_0}E_{t_0})}\, \sinh{[(N+1)\beta_{t_0}E_{t_0}]} }\,.\label{eq:h}
\end{align}
In Eqs.~\eqref{eq:probgen_indist} and \eqref{eq:probgen_dist}, the common protocol-dependent, and $N$-, $\beta_{t_0}$-independent parts of the first three terms given by $\vert d_i(t_0) \vert^2$ and $\vert \tilde{c}^{\pm}_i(t_0) \vert^2$ are positive but the third term (on the second line of the equations) proportional to $\Re [d_i(0)d_i^*(T/2)]$ is not necessarily positive. As discussed in the main paper, for $\Delta = 0$,  $\cos{\theta_t} = 0$ at all $t$ and the non-positive term vanishes. In this case, one can directly compare the probabilities for the indistinguishable and distinguishable cases in Eqs.~\eqref{eq:probgen_indist} and \eqref{eq:probgen_dist}, by examining the relative size of the $N$- and $\beta_{t_0}$-dependent factors. To this end, we use the following relation that holds for any $N>1$ and $x>0$:
\begin{align}
\left[\frac{N}{2}\left(\frac{N}{2}+1 \right)-F_{\pm}(N,x) \right] \geq \left[\frac{N}{2}(1\mp\tanh{(x)}) \right]. \label{eq:cond2_supp}
\end{align}
This leads to one of the central results given in the main paper: when $\Delta=0$, for any form of the coupling protocol $g_{C}(t)$, $N>1$, system Hamiltonian and value of $\beta_c$ and $\beta_h$, one finds that $p_i^{\mr{indist}}>p_i^{\mr{dist}}$, i.e., the internal energy change done in the indistinguishable case is larger than the distinguishable case. 

Let us now consider $\Delta \neq 0$. While in this case the enhancement of internal energy change for indistinguishable engines is not guaranteed in general, in what follows we identify conditions under which enhancement can be expected. We first note that the inequality \eqnref{eq:cond1_supp} allows us to compare the first term of Eqs.~\eqref{eq:probgen_indist} and \eqref{eq:probgen_dist} proportional to $\vert d_i(t_0)\vert^2$ and see it is larger in the indistinguishable case. Secondly, the $N$- and $\beta_{t_0}$-dependent positive factors multiplying $2\Re [d_i(0)d_i^*(T/2)]$ satisfy
\begin{align}
4 h(N,\beta_{c}E_{0})\, h(N,\beta_{h}E_{T/2}) \geq N^2 \tanh{(\beta_{c}E_{0})} \tanh{(\beta_{h}E_{T/2})} \label{eq:cond3},
\end{align}
for any $N>1$, $\beta_cE_0$, and $\beta_h E_{T/2}$, with the equality holding for $N=1$. 
Thus, from the above discussion together with the previous considerations leading up to \eqnref{eq:cond2_supp}, it follows that $p_i^{\mr{indist}}>p_i^{\mr{dist}}$ which enhances the average internal energy change, for coupling protocols and external system Hamiltonians that satisfy the constraint $\Re [ d_i(0) d_i^*(T/2)] \geq 0$. Note that $h(N, \beta_{c}E_{0})\, h(N,\beta_{h}E_{T/2})$ is an increasing function of $N$ with magnitude comparable to the functions on the left-hand side of the inequalities in Eqs. \eqref{eq:cond1_supp} and \eqref{eq:cond2_supp} for small $N$. Thus, for small $N$, the magnitude of the negative term when $\Re [ d_i(0) d_i^*(T/2) ] < 0$ is off-set by the positive terms in \eqnref{eq:probgen_indist} and  enhancement persists. In other words, for small enough $N$,  enhancement is generally present, regardless of  the form of the coupling function, the external system Hamiltonian, and the temperatures of the baths, even when $\Delta \neq 0$.

\begin{figure*}[t]
	\centering
	\includegraphics[width= 0.8\columnwidth]{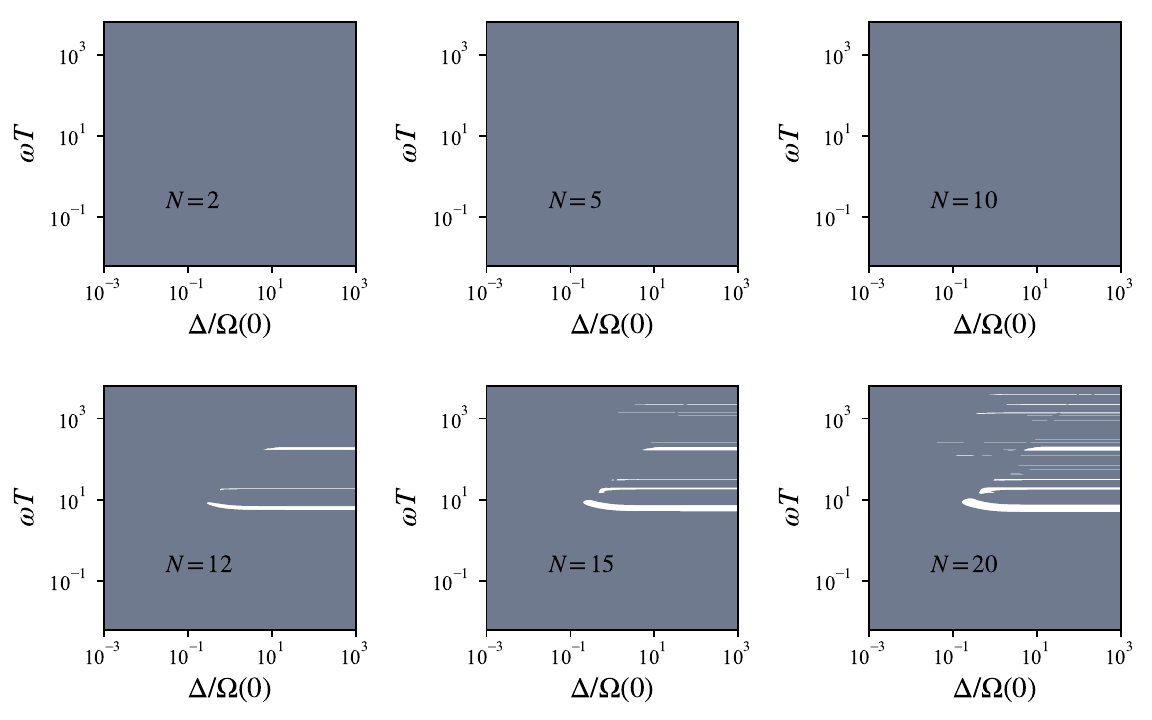}
	\caption{
		{\bf Mapping out regions of enhancement in parameter space.} Binary contour plot of regions for different values of the number of engines, as a function of $\Delta/\Omega(0)$ and $\omega T$, with enhancement $\mathcal{E} > 1$ (grey) and no enhancement $\mathcal{E} \le 1$ (white) calculated via the analytical formula presented in Eqs.~\eqref{eq:probgen_indist} and \eqref{eq:probgen_indist}. The remaining parameters are chosen as in the main paper: $g = 0.01$, $\delta_t = 0.9$, $\alpha = 2142/T$, $t_{\rm on} = (1-\delta_t)T/4$, and $t_{\rm off} = t_{\rm on}+\delta_t T/2$. Engine parameters are $v = 0.1 \Omega(0)^2$, $T = 20/\Omega(0)$, $\beta_c E_0 = 2$, and $\beta_h E_{T/2} = 1/4$.}
	\label{fig:figapp1}
\end{figure*}

We will now demonstrate that we can, without much fine tuning, identify parameter regimes where enhancement is possible for a given choice of the external system Hamiltonian and smooth form for the coupling function. Let us choose for the external system, a harmonic oscillator (HO) of frequency $\omega$ coupled to the engines. For the coupling, we consider a  function which is constant for most of the duration of the work stroke and turns on and off smoothly. A suitable choice is the one used in our previous work \cite{watanabe17suppl} \begin{align}
g_C(t) =  \frac{g}{\delta_t T} \sum_{n=0}^1 \left\{ \tanh{\left[\alpha \left( t - t_{\rm on} - \frac{nT}{2} \right) \right]} - \tanh{\left[\alpha \left( t - t_{\rm off} - \frac{nT}{2} \right) \right]} \right\}\,,\label{eq:gc_cont}
\end{align}
where $\alpha$ is the switching rate and $\delta_t$ $(0<\delta_t<1)$ is the ratio of the duration in which the coupling is turned on from $t_{\rm on} + nT/2$ to $t_{\rm off} + nT/2$ $(n=0$ and $1)$ in the duration of the half of the cycle for $T/2$: $t_{\rm on} = (1-\delta_t)T/4$ and $t_{\rm off} = t_{\rm on}+\delta_t T/2$. We note that this is the coupling form used in the numerical calculations generating the results in Figs.~3 and 4 of the main paper. Let us, for the sake of concreteness, assume that the engine compression and expansion strokes are represented by a linear protocol with velocity $v$ as described in the main paper. We would now like to identify the parameter choices for which $\Re [ d_i(0) d_i^*(T/2) ] > 0$ holds and hence results in enhancement. First, we notice that for the HO external system ${_S}\wich{i}{V_S^{(I)}(t)}{0}_S = e^{i\omega t} \delta_{i,1}$, i.e. in the weak coupling, perturbative treatment only the first excited state is occupied. Secondly, for simplicity we take the coupling function to have a fast switching rate with $\delta_t \approx 1$ allowing the approximation $g_C(t)\approx g/T$. Thus, we have
\begin{align}
d_1(t_0) = -\int_{t_0}^{t_0+T/2} dt\, \frac{g}{T}\, e^{i\omega t} \frac{\Delta}{\Omega(t)^2+\Delta^2}\,. \label{eq:dit0}
\end{align}
When the external system is an oscillator with small frequency $\omega T  \ll 1$, we can ignore the oscillating phase factor in the integral and find
\begin{align*}
d_1(t_0) \approx -\frac{g \Delta}{vT}\, \log \left[ \frac{\sec{\theta_{T/2}}+\tan{\theta_{T/2}}}{\sec{\theta_0} + \tan{\theta_0}}\right],
\end{align*}
leading to
\begin{align}
\Re[d_1(0) d_1^*(T/2)] = \frac{g^2 \Delta^2}{v^2 T^2} \left( \log \left[ \frac{\sec \theta_{T/2}+\tan{\theta_{T/2}}}{\sec{\theta_0} + \tan{\theta_0}}\right] \right )^2 \geq 0 \label{eq:smallw_badfac}.
\end{align}
Thus, in this limit of small oscillator frequency a quantum enhancement is always present. On the other hand when $\omega T \gg 1$, 
\begin{align*}
d_1(t_0) \approx \frac{-g\Delta}{i\omega T} \left[\frac{e^{i \omega (t_0+T/2)}}{\sqrt{\Omega(t_0+T/2)^2+\Delta^2}} - \frac{e^{i \omega t_0}}{\sqrt{\Omega(t_0)^2+\Delta^2}}\right],
\end{align*}
and
\begin{align}
\Re[d_i(0) d_i^*(T/2)] \approx \frac{g^2 \Delta^2}{\omega^2 T^2} \left[\frac{2\cos{(\omega T/2)}}{\sqrt{[\Omega(0)^2+\Delta^2][\Omega(T/2)^2+ \Delta^2]}} -\frac{\cos{(\omega T)}}{\Omega(0)^2+\Delta^2} - \frac{1}{\Omega(T/2)^2+\Delta^2} \right ] \label{eq:lgw_badfac},
\end{align}				
with $\Omega(T/2) = \Omega(0)+vT/2$. The presence of the oscillating factors in the above expression can lead to $\Re[d_i(0) d_i^*(T/2)]<0$. While the dependence on $\Delta$, $\Omega(0)$, and $v$ implied by Eq.~\eqref{eq:lgw_badfac} is complicated, we can immediately see that when $\Delta \ll \{\Omega(0),\omega\}$ the magnitude of the term is suppressed. Thus even if  $\Re[d_i(0) d_i^*(T/2)]<0$, in this limit this term will not be able to suppress the other positive terms in $p_i^{\mr{indist}}$. Thus we can anticipate that in the small $\Delta$ regime, in agreement with the idea of continuity with the $\Delta = 0$ result,  enhancement is exhibited. In the other extreme, when $\Omega(0) = 0$ and $\Delta \gg v T$ \eqnref{eq:lgw_badfac} reduces to
\begin{align*}
\Re[d_i(0) d_i^*(T/2)] \approx \frac{g^2}{\omega^2 T^2} \left[2 \cos{(\omega T/2)}-\cos{(\omega T)} - 1\right].
\end{align*}
In this regime,  varying $\omega T > \pi$ thus unveils the regions in which enhancement appears and disappears as a result of  the oscillatory nature of the sign of $\Re[d_i(0) d_i^*(T/2)]$. To summarize and support the above discussion,  \figref{fig:figapp1} shows contour plots, as a function of $\Delta/\Omega(0)$ and $\omega T$, of the regions where enhancement is present (grey) or absent (white) for different fixed values of $2 \le N \le 20$. Here, the expressions \eqref{eq:probgen_indist} and \eqref{eq:probgen_dist} are used to calculate the average internal energy change in the indistinguishable and distinguishable case. The main features are in agreement with the analytical arguments above, namely, enhancement is present at small $N$ independent of other parameter choice (for all the parameter space region explored).  Further, regions with no enhancement appear for $\omega T > \pi$  at large enough $N$. Such regions also shrink and vanish as $\Delta/\Omega(0) \rightarrow 0$.

\subsection{Work content of the energy delivered to the system}

\begin{figure}[t]
	\centering
	\includegraphics[width= \columnwidth]{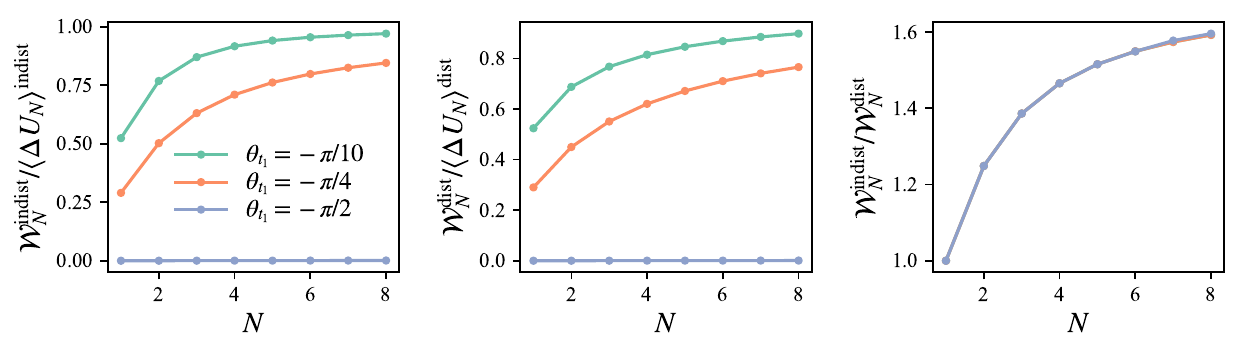}
	\caption{Ratio of the ergotropy $\mathcal{W}_N$ to the average energy of the harmonic oscillator external system state for $N$ indistinguishable bosonic (left) and distinguishable (middle) work resources. Ratio of ergotropies for the indistinguishable to distinguishable case (right). The parameters and settings are the same as in Fig.~2 [impulse-type coupling] of the main paper. Note that, although the ergotropy for $\theta_{t_1}=-\pi/2$ shown in the left and the middle panels are small, it is nonzero. In the right panel, results for all the three values of $\theta_{t_1}$ almost overlap.}
	\label{fig:ergo_fig2ms}
\end{figure}
\begin{figure}[tb]
	\centering
	\includegraphics[width=\columnwidth]{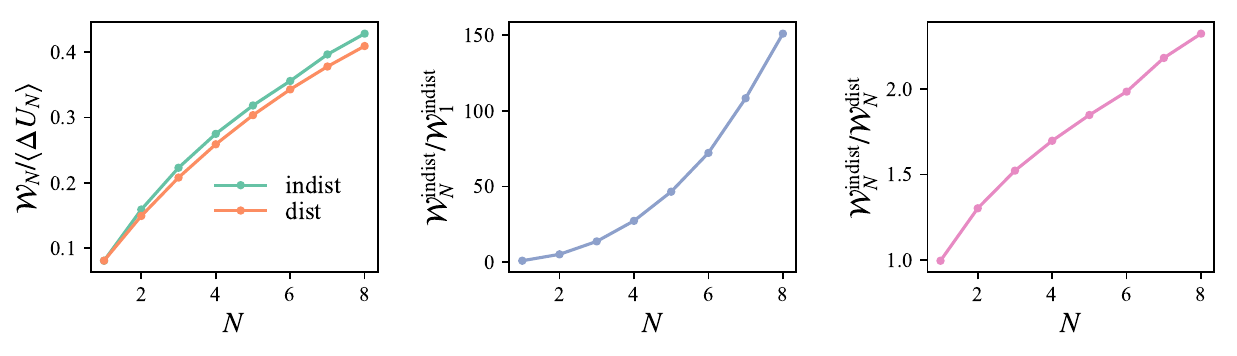}
	\caption{Ratio of the ergotropy $\mathcal{W}_N$ to the average energy of the harmonic oscillator external system state for $N$ indistinguishable bosonic and distinguishable work resources (left). Ergotropy for $N$ work resources in units of the ergotropy for $1$ resource (middle).  Ratio of ergotropies for the indistinguishable to distinguishable case (right). The parameters and settings are the same as in Fig.~3 [continuous nonperturbative coupling] of the main paper.}
	\label{fig:ergo_fig3ms}
\end{figure}

The increase in the energy content of the system $\Delta U$ is a consequence of the combined action of the engines which convert heat flowing from the hot to the cold reservoir into work, and the time dependence of the coupling constant $g_C$ between the engines and the system $S$. In what follows, we provide one method to isolate the work contribution due to the engines alone and, moreover, show that this exhibits collective enhancement for the parameter regimes considered in the main paper.

As a first task, we break the internal energy change (since the external system $S$ initially starts from the ground state with zero initial energy in our setup, this is a property purely of the final state) into ergotropy $\mathcal{W}$ \cite{lenard78suppl,pusz78suppl,francica17suppl}, which is a measure of the maximum work capacity of the final state of the system with respect to its (uncoupled) Hamiltonian, and the average energy $E_{\mathrm{pas}}$ of a passive state that cannot deliver work under cyclic unitary processes, i.e.,
\begin{align}
\avg{\Delta U} = \mathcal{W} + E_{\mathrm{pas}} \label{eq:ergbreakup}.
\end{align}
Thus, ergotropy provides an important energetic measure and quantifies how much of the energy exchanged with the system can be utilized in a later operation with the external system acting as a battery and we will take this as the quantifier of the total work done on the system. In order to identify the fraction of this work contributed purely by turning the coupling to the system on and off, we need to separate this part from the contribution of the action of the engines through the dynamics of their operator $V_R^{(I)}(t)$ in the coupling term $g_C(t)\, V_R \otimes V_S$. To evaluate the former contribution, we focus on the external system $S$ part and consider a time-dependent Hamiltonian $H_{\rm driv}(t) \equiv H_S + g_C(t)\, \bar{V}_{R,\, t} V_S$, which is externally driven solely by the time-dependent parameter $g_C(t)$. Here, $\bar{V}_{R,\, t}$ is a piecewise-constant factor during each work stroke given by the time average of the work resource coupling operator $V_R$ as
\begin{align}
\bar{V}_{R,\, t} \equiv \begin{cases} 
\displaystyle \frac{2}{T} \int_{0}^{T/2} dt'\, \avg{V_R^{(I)}(t')}_{\rho_0^R} \,\,&\mathrm{for}\,\, 0 < t < \dfrac{T}{2}\,,\\[.75em]
\displaystyle \frac{2}{T}  \int_{T/2}^{T} dt'\, \avg{V_R^{(I)}(t')}_{\rho_{T/2}^R} \,\,&\mathrm{for}\,\, \dfrac{T}{2} < t < T\,.
\end{cases}
\label{eq:suppVbardriv}
\end{align}
Here, the driving acts on the system via the same operator $V_S$ as in Eq.~(2) of the main paper and the averages inside the integrals in Eq.~\eqref{eq:suppVbardriv} are taken with respect to the state of the engines in thermal equilibrium with the cold ($\rho_0^R$, compression stroke, $0<t<T/2$) and hot ($\rho_{T/2}^R$, expansion stroke, $T/2<t<T$) baths. We identify the average energy change $E_{\mathrm{driv}} \equiv \avg{H_{\rm driv}(T)} - \avg{H_{\rm driv}(0)}$ of the external system part caused by the unitary process by the time-dependent Hamiltonian $H_{\rm driv}(t)$ as the energy change due to the time dependence of $g_C(t)$. This can be a good estimate provided the time dependence of $V_R^{(I)}(t)$ is sufficiently weak. Since this unitary process is cyclic due to the form of the coupling $g_C(t)$, the energy change (starting from the passive ground state of the system) will also be equal to the ergotropy $\mathcal{W}_{\mathrm{driv}}$ of the external system's final state, i.e., $E_{\mathrm{driv}} = \mathcal{W}_{\mathrm{driv}}$. With this, we propose that the difference in ergotropies between the engine-driven case and time-dependent Hamiltonian case, $\Delta \mathcal{W} \equiv \mathcal{W}-\mathcal{W}_{\mathrm{driv}}$, be taken as the proxy of the work delivered to the external system from the engine action alone. We will only consider the continuous coupling case for such estimates of $\Delta \mathcal{W}$ since the impulse-type coupling has a nonzero value only at an instance of time and this makes difficult to separate the contribution of the energy injection due to the time-dependent coupling constant from that of the action of the engines. We next present results for the behavior of both $\mathcal{W}$ and $\Delta \mathcal{W}$ for the settings considered in the main paper.

\begin{figure*}
	\centering
	\includegraphics[width = 0.6\textwidth]{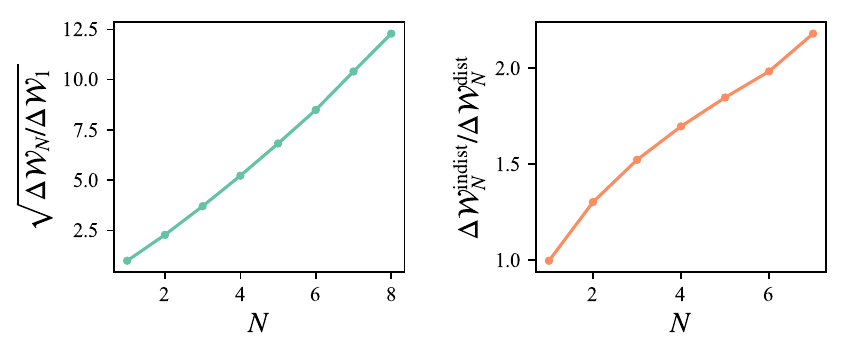}
	\caption{Left: Ratio of the newly defined estimate of work output for $N$ indistinguishable engines to $1$ engine. Right: Ratio of the newly defined estimate of work output for $N$ indistinguishable engines to $N$ distinguishable engines. Parameters are same as in Fig.~3 of the main paper.}
	\label{fig:figS4}
\end{figure*}
\begin{figure*}
	\centering
	\includegraphics[width = 0.6\textwidth]{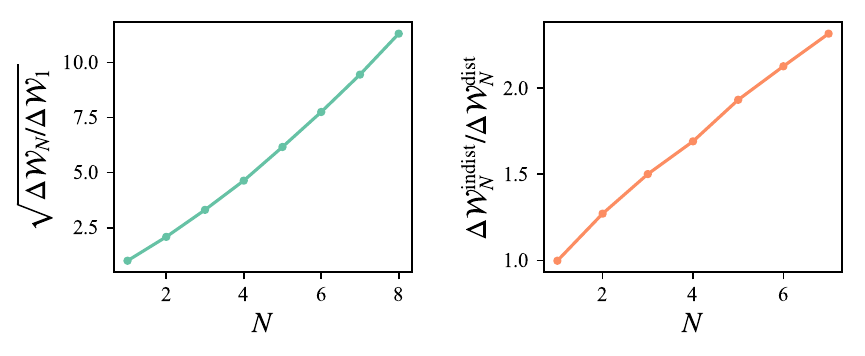}
	\caption{Left: Ratio of the newly defined estimate of work output for $N$ indistinguishable engines to $1$ engine. Right: Ratio of the newly defined estimate of work output for $N$ indistinguishable engines to $N$ distinguishable engines. $\Delta = 0.1 \Omega(0)$, and the rest of the parameters are same as in Fig.~3 of the main paper.}
	\label{fig:figS5}
\end{figure*}

First, we have performed a numerical calculation of the ergotropy of the external system taken as a harmonic oscillator coupled to $N$ distinguishable and indistinguishable bosonic work resources for the situations presented in Figs.~2 and 3 of the main paper. We show the results in Figs.~\ref{fig:ergo_fig2ms} and \ref{fig:ergo_fig3ms}. From this calculation, it is clear that both impulse-type and continuous coupling of the external system to the work resources results in the external system accumulating ergotropy and not merely passive energy. Moreover, in the both cases, the ergotropy displays enhanced values for the indistinguishable bosonic engines case as opposed to the distinguishable one. Additionally, we note that, in the impulse-type coupling case, the ergotropy strongly depends on the value of the variable $\theta_{t_1}$ (decided by the time of the kick) but the enhancement looks to be almost independent of this variable.

Finally, for the continuous coupling case, we consider $\Delta \mathcal{W}$, the estimate of the work output due to the engine action we introduced. In Fig.~\ref{fig:figS4}, we present the results from a numerical calculation of $\Delta \mathcal{W}$ for the parameters in Fig.~3 (nonperturbative continuous coupling) of the main paper. Here we clearly find that $\Delta \mathcal{W}$ is positive, i.e., the engine contributes to the work content of the system and, moreover, this quantity also exhibits collective enhancement for the parameters considered in the main paper. We have also included another case with $\Delta \neq 0$ shown in Fig.~\ref{fig:figS5}, and see the similar behavior as in Fig.~\ref{fig:figS4}.
We have checked that, as far as $\Delta/\Omega(0)$ is relatively small, this behavior (positive $\Delta \mathcal{W}$ and enhancement) holds for continuous coupling between the engines and the system provided $g$ is sufficiently large or $\omega \sim \Omega(0)$, both of which allow efficient energy transfer from the engines to the external system for $\Delta/\Omega(0) \ll 1$.


\subsection{Multiple fermionic engines trapped in a harmonic oscillator potential}

\begin{figure}[t]
\centering
  \includegraphics[width = 0.6\textwidth]{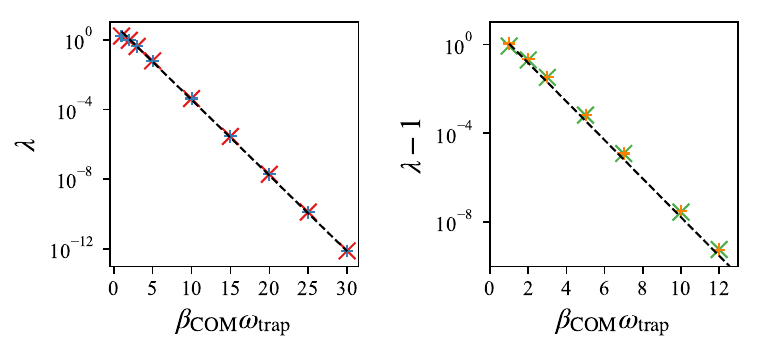}
  \caption{
    {Performance of fermionic engines at nonzero COM temperature $\beta_{\rm COM}^{-1}$.}
The left panel shows the internal energy increase of the system $S$ for even $N$, with $N=2$ (red) and $N=4$ (blue) obtained by numerical simulations. They are well reproduced by the analytical expression $\lambda = 8 \exp{(-\beta_{\rm COM} \omega_{\rm trap})}$ (dashed line). Odd values of $N$ are considered in the right panel, with $N=3$ (green) and $N=5$ (orange), in good agreement with $\lambda -1 = 8 \exp{(-2\beta_{\rm COM} \omega_{\rm trap})}$ (dashed line). Parameters are $\Delta = 1$, $v= 0.5/\Delta^2$, $T=20/\Delta$, $\beta_c E_0 = 1$, $\beta_h E_{T/2} = 1/8$, $\omega = 2\pi \times 0.05/T$, $g = 0.5$, $\delta_t = 0.98$, and $\alpha = 2000/T$.
}
\label{fig:fermi}
\end{figure}

We discuss the case in which each engine is made of an identical non-interacting two-level fermionic atom. In this case, the average internal energy change $\avg{\Delta U}_N$ of the system $S$ displays a pronounced even-odd dependence with respect to $N$. When the temperature $\beta_{\rm COM}^{-1}$ of the center of mass (COM) degrees of freedom is zero, $\avg{\Delta U}_N=0$ for even $N$, and $\avg{\Delta U}_N=\avg{\Delta U}_1$ for odd $N$. This is because the atoms inside the Fermi sphere, in which both the ground and excited internal levels are fully occupied, cannot contribute to internal energy change because of the Pauli blocking: In the case of odd $N$, there is one fermion on the top of the Fermi surface, and this is the only atom which can change the internal states in the engine operation. By contrast, there is no such atom for even $N$. Thus in this regime, the internal energy change of the external system due to $N$ work resources is upper-bounded by $\avg{\Delta U}_1$ and hence in the limit of perturbative coupling, we can anticipate that this quantity will be diminished with respect to the internal energy change for distinguishable engines which would scale linearly with $N$. At nonzero $\beta_{\rm COM}^{-1}$, $\avg{\Delta U}_N$ takes non-trivial values due to the contribution of atoms near the Fermi surface, where the COM levels are only partially occupied. Moreover, in the regime where $\beta_{\rm COM}^{-1}$ is much smaller than the energy difference $\omega_{\rm trap}$ between the highest occupied and lowest unoccupied trap levels of the atoms, the ratio $\lambda \equiv \avg{\Delta U}_N/\avg{\Delta U}_1$ for an arbitrary trap in general can be well characterized by only $\beta_{\rm COM} \omega_{\rm trap}$. In Fig.~\ref{fig:fermi} we show $\lambda$ numerically calculated for multiple fermionic engines in the Otto cycle for several values of $N$. In this example, we assume that atoms are trapped in a HO potential with frequency $\omega_{\rm trap}$ and another HO with frequency $\omega$ is taken as an external system $S$. At higher COM temperatures $\beta_{\rm COM}^{-1}$ for which $\beta_{\rm COM}\omega_{\rm trap} \alt 1$, $\avg{\Delta U}_N$ depends on details of the system such as the spectrum of the COM degrees of freedom determined by the shape of the trapping potential and the Fermi energy. A detailed study of this dependence is beyond the scope of the present work and will be taken up in the future.

We now derive an analytical estimate for the ratio $\lambda$ when $\beta_{\rm COM}^{-1} \omega_{\rm trap} \gg 1$. We will obtain this estimate by considering isolated engines without coupling to an external system as we show below. The Hamiltonian $H_E(t)$ of the engines consists of that of the COM degrees of freedom $H_{\rm COM}$ and that of the internal degrees of freedom $H_{\rm in}(t)$, $H_E(t) = H_{\rm COM} + H_{\rm in}(t)$, with
\begin{align}
  H_{\rm COM} &= \omega_{\rm trap} \sum_{l} \left( l + \frac{1}{2} \right) \left( a_l^\dagger a_l + b_l^\dagger b_l \right)\,,\\
  H_{\rm in}(t) &= \sum_{l} \left[ \Omega(t)\, (a_l^\dagger a_l - b_l^\dagger b_l) + \Delta\, (a_l^\dagger b_l + b_l^\dagger a_l) \right]\,.
\end{align}
Here, $\omega_{\rm trap}$ is the frequency of the HO trapping potential whose levels are labeled by $l$. The operators $a_l^\dagger$ ($a_l$) and $b_l^\dagger$ ($b_l$) are creation (annihilation) operators of a spin-up and spin-down atom at $l$th level of the trap, respectively, satisfying the fermionic anti-commutation relations $\{a_l,\, a_{l'}^\dagger\}=\{b_l,\, b_{l'}^\dagger\} = \delta_{l,\, l'}$ and $\{a_l,\, a_{l'}\} = \{a_l^\dagger,\, a_{l'}^\dagger\} = \{b_l,\, b_{l'}\} = \{b_l^\dagger,\, b_{l'}^\dagger\} = 0$ with $\{A,\, B\} \equiv AB + BA$ being the anti-commutator. The distribution of the population $n_l \equiv a_l^\dagger a_l + b_l^\dagger b_l$ of the trap levels is conserved, which is set by the COM temperature $\beta_{\rm COM}^{-1}$ and follows the canonical distribution $\propto \exp{(-\beta_{\rm COM} H_{\rm COM})}$.

Using the internal degrees of freedom of the two-level fermionic atoms described by the above Hamiltonian, we consider the quantum heat engines in the Otto cycle. The average of work done by $N$ isolated engines without outcoupling is
\begin{align}
  \avg{w}_N = (\epsilon_h - \epsilon_c) (\tanh{\beta_c\epsilon_c} - \tanh{\beta_h\epsilon_h})\,\, f_{N}(\beta_{\rm COM}\omega_{\rm trap})\,,\label{eq:w_N}
\end{align}
with $\pm \epsilon_h \equiv \pm \sqrt{\Delta^2 + \Omega_h^2}$ and $\pm \epsilon_c \equiv \pm \sqrt{\Delta^2 + \Omega_c^2}$ being the energy eigenvalues of the internal degrees of freedom in the hot and cold isochore process, respectively. Here, $f_{N}$ is a function of $\beta_{\rm COM}\omega_{\rm trap}$ whose detailed form varies with $N$. Since $f_{N=1}=1$, the ratio $\lambda$ is
\begin{align}
  \lambda = f_{N}(\beta_{\rm COM}\omega_{\rm trap})\,,
\end{align}
which is independent of the temperatures of the heat baths and the spectrum of the internal degrees of freedom.
Although the exact form of $f_{N}$ non-trivially depends on $N$ in general, in the low COM temperature limit of $\beta_{\rm COM}\omega_{\rm trap} \gg 1$ the form of $f_{N}$ differs only by the parity of $N$ as
\begin{equation}
  f_{N}(\beta_{\rm COM}\omega_{\rm trap}) \simeq \left\{
  \begin{array}{ll}
    8 e^{-\beta_{\rm COM}\omega_{\rm trap}}\quad & \mbox{(even $N$)}\,,\medskip\\
    1 + 8 e^{-2\beta_{\rm COM}\omega_{\rm trap}}\quad & \mbox{(odd $N$)}\,.
  \end{array}
  \right.\label{eq:lambda_N}
\end{equation}
Interestingly, though Eq.~(\ref{eq:lambda_N}) is obtained for the isolated engines without outcoupling, we find that the numerical result (symbols) for $\lambda$ in Fig.~\eqref{fig:fermi} is well fitted by the above analytical formula (dashed lines). Finally, we add that even though Eq.~(\ref{eq:lambda_N}) is obtained for the HO trapping potential, it is applicable for any type of the trapping potential in the low temperature limit, $\beta_{\rm COM}\omega_{\rm trap} \gg 1$, by replacing $\omega_{\rm trap}$ by the energy difference between the highest occupied and the lowest unoccupied levels of the trap at zero COM temperature.


\begin{thebibliography}{99}
\bibitem{note:general} In this framework, generalization for the coupling Hamiltonian $H_C(t) =  \sum_l g_{C,\, l}(t)\, V_{R,\, l} \otimes V_{S,\, l}$  ($l$ labels the work resource) is straightforward.
\bibitem{note:autocorrfunc} Note that for $T/2\leq \{t,t^{\prime}\}\leq T$, due to the thermalization step at $t=T/2$, in \eqnref{eq:probcc_gen} we take the autocorrelation function with respect to the state $\rho_{T/2}^R$.
\bibitem{note:adiabaticity} For $\Delta\equiv 0$, the above results are accurate even in the diabatic case since the exact time evolution operator and the adiabatic time evolution operator of the engine coincide for this case since $H_E$ is diagonal throughout.
\bibitem{note:dicke} It should be noted that the $N^2$ scaling is obtained for the thermal initial state of the engines, which is strikingly different from the Dicke superradiance \cite{dicke54,garraway11}.
\end{thebibliography}
\end{document}